# Lien entre fermentations microbiennes *in vitro* et performances *in vivo* chez les porcs sélectionnés sur leur consommation résiduelle

Olivier ZEMB (1), Lauren JOUARON (1), Jordi ESTELLE (2), Anais CAZALS (2), Caroline ACHARD (3), Marion SCHIAVONE (3), Tiffany PAGE (1), Laurent CAUQUIL (1), Carole BANNELIER (1), Amir ALIAKBARI (1), Yvon BILLON (4), Yves FARIZON (1), Hélène GILBERT (1)

(1)GenPhySE, Université de Toulouse, INRAE, ENVT, 31326, Castanet-Tolosan, France
(2)GABI, INRA, AgroParisTech, Université Paris-Saclay, 78350 Jouy-en-Josas, France
(3)Lallemand SAS, 19, rue des briquetiers, BP 59, 31702 Blagnac, France
(4)GenESI, INRAE, 17700 Surgères, France
Olivier.zemb@inrae.fr

*Lien entre fermentations microbiennes in vitro et performances in vivo chez les porcs sélectionnés sur leur consommation résiduelle*

*Cette étude a analysée les différences de microbiote intestinal entre des porcs de lignées efficaces, cad à la consommation moyenne journalière résiduelle faible (CMJR-) par rapport à des porcs non-efficaces (CMJR+). Le séquençage a montré un lien entre l'efficacité et certaines voies métaboliques, notamment pour la voie de la biosynthèse des acides aminés et la voie de synthèse de l'ARNt-aminoacyl chez les CMJR+ qui est associée à une production accrue de propionate. Cependant les analyses de profils de fermentation in vitro montrent que les porcs CMJR–, de la lignée plus efficace, produisent plus d'acétate (+15%) et de propionate (+56%) à partir des fibres insolubles (IF) récupérées après simulation de digestion par la partie haute du tube digestif. Le valérate est également plus fréquemment abondant chez les CMJR– (P < 0,01). L'analyse par séquençage 16S des microbes responsables de la fermentation suggère que le propionate obtenu à partir de la fraction de l'aliment qui est indigestible par l'hôte est principalement produit par Prevotella et Lactobacillus. Cette production est fortement corrélée à l'épaisseur du lard dorsal chez les porcs de la lignée CMJR– (Corrélation de Spearman = 0,8), tandis que chez les porcs CMJR+, une corrélation modérée existe entre la production de butyrate et l'indice de consommation (Corrélation de Spearman = 0,44). Ces résultats montrent que la production de propionate est corrélée au métabolisme adipeux, suggérant que l'activation du récepteur GPR43 par le propionate pourrait jouer un rôle physiologique dans les cellules adipeuses des porcs CMJR–. Ces observations mettent en évidence des différences fonctionnelles importantes entre les microbiotes des porcs CMJR+ et CMJR– ainsi qu'une variabilité au sein des porcs génétiquement plus efficaces, qui pourrait être exploitée pour améliorer les performances.*

*Relation between in vitro microbial fermentations and in vivo performance in pigs selected for their residual feed intake*

*Bioinformatic analysis of microbiota revealed that certain metabolic pathways are associated with low- and high- residual feed intake (HRFI and LRFI), such as the amino-acid biosynthesis pathway and the tRNA-aminoacyl synthesis pathway. The latter is associated with increased propionate production. Yet, in vitro fermentation-profile analyses revealed that LRFI pigs, from the most efficient genetic line, produced more acetate (+15%) and propionate (+56%) from the insoluble fraction (IF) containing the insoluble dietary fibre recovered after simulation of upper gastrointestinal digestion. Valerate was also more frequently abundant in LRFI pigs (P < 0.01). 16S sequencing analysis of the microbes responsible for fermentation suggested that propionate obtained from the fraction of feed that is indigestible by the host is produced mainly by Prevotella and Lactobacillus. This production was strongly correlated with backfat thickness in LRFI pigs (Spearman's correlation = 0.80), while a moderate correlation existed between butyrate production and feed efficiency in HRFI pigs (Spearman's correlation = 0.44). These results revealed that propionate production is related to fat metabolism, suggesting that GPR43 receptor activation by propionate could play a physiological role in adipose cells in RFI- pigs. These observations highlight significant functional differences between the microbiota of HRFI and LRFI pigs, as well as variability within more efficient pigs that could be exploited to improve performance.*

**INTRODUCTION**

Après le broyage par mastication, la digestion de l'aliment se poursuit par l'action de la pepsine dans l'estomac et à la pancréatine dans l'intestin grêle. Ce dernier est alors en mesure d'absorber une grande partie des molécules solubilisées, tandis que la partie insoluble de l'aliment non digérée en amont parvient jusqu'au colon où les éléments dégradables (fibres solubles, amidon résistant…) sont fermentés par les microbes intestinaux en acides gras à chaine courte (Sonnenburg, Sonnenburg, 2014), dont la concentration atteint typiquement entre 50 et 200 mM dans le colon chez l'homme et jouent un rôle en tant que source d'énergie pour l'épithélium intestinal et pour la néoglucogénèse dans le foie (Louis, Flint, 2017).

La comparaison de lignées de porcs divergentes en termes de consommation moyenne journalière résiduelle (CMJR) a montré que les porcs à CMJR réduite (sélectionnés pour être plus efficaces, CMJR–), avaient des microbiotes fécaux différents des porcs CMJR+ (moins efficaces) : ces différences concernent notamment les genres microbiens Dialister, Clostridium_sensu_stricto_1, Prevotella_7, Terrisporobacter (Aliakbari et al., 2021). Dans cette étude, nous évaluons sur certains de ces animaux les différences de fonctionnement du microbiote fécal en utilisant l'inférence fonctionnelle et des fermentations in vitro de la fraction insoluble qui parvient jusqu'au colon. Les résultats obtenus sont comparés aux performances des animaux.

**1. MATERIELS ET METHODES**

**1.1. Animaux, mesures et prélèvement des échantillons**

Les échantillons de 24 porcs de chaque lignée CMJR- et CMJR+ décrites précédemment (Gilbert, 2015), élevés dans les trois mêmes groupes de contemporains en 2016 et 2017 sur l'élevage expérimental GenESI (Surgères, France, https://doi.org/10.15454/1.5572415481185847E12), ont été utilisés pour cette étude. Ces porcs ont fait l'objet de mesures de performances entre 10 semaines d'âge et l'abattage (poids cible de 115 kg de poids vif) : des poids vifs ont été mesurés en début et fin de test, l'ingéré journalier individuel était enregistré par un distributeur automatique de concentré (DAC ACEMA 64, ACEMO Skiold, Pontivy, France), permettant de calculer la consommation moyenne journalière et l'indice de consommation individuel (IC) pour chaque porc. Des mesures d'épaisseurs de lard (ELD) par ultra-sons ont été réalisées en fin de test en six points du dos et moyennée pour obtenir une estimation de la composition corporelle individuelle. Les valeurs individuelles de CMJR ont été calculées (Aliakbari et al., 2021). Les porcs avaient un accès libre à l'eau et à un aliment granulé commercial de type croissance, contenant en moyenne 10 MJ d'énergie nette et 160 g de protéines par kg.

Les 48 échantillons de fèces ont été prélevés à 15 semaines d'âge, congelés à -20°C immédiatement, et transportés congelés (Aliakbari et al., 2022). L'aliment distribué en engraissement a été prélevé à la même date, et envoyé à température ambiante au laboratoire.

**1.2. Inférence fonctionnelle par PICRUST2**

Les données de séquençage du microbiote fécal de 588 animaux publiées précédemment (Aliakbari et al., 2021) ont été traitées par PICRSUT2 avec les paramètres par défaut (Langille et al., 2013), pour prédire les fonctions microbiennes présentes dans les échantillons ainsi collectés.

**1.3. Préparation des substrats à fermenter**

Différents types de substrats de fermentations ont été testés : l'aliment utilisé dans l'élevage (Aliakbari et al., 2021) broyé à 0,5 mm sur la machine Pulverisette 14 Fritsch, l'amidon de maïs

(Sigma-Aldrich, S4126), les beta-glucanes d'avoine (SoMatEM GmbH Freschen) et la fraction de l'aliment indigestible par l'hôte, pour laquelle une première digestion in vitro a été effectuée pour simuler la digestion haute (Jha et al., 2012). Brièvement, 10 g d'aliment broyé sont placés dans du tampon PBS à pH 6 à 38°C à 50 rpm pendant 2 h avec 20 mg/mL de pepsine et 200 mL de HCl qui permet de faire descendre le pH à 4 pour simuler la digestion stomacale. Puis le pH est rétabli à 7 avec environ 100 mL de NaOH 0,6 M et la suspension est incubée à 38°C, 50 rpm pendant 4 h avec de la pancréatine à 100 mg/mL pour simuler la digestion duodénale. La suspension est ensuite divisée en une fraction soluble (SF) et une fraction insoluble (IF) par une filtration sur un tamis métallique de 50 microns. La fraction insoluble est lavée à l'éthanol 95%, puis mise à sécher pendant 48 h à 60°C tandis que la fraction soluble est mélangée à de l'éthanol chauffé à 60°C avec un ratio volumétrique 2:1 (Mou et al., 2020). Après une incubation de 2 h permettant la précipitation des protéines, le mélange est filtré, lavé à l'éthanol et séché à 60°C pendant 48 h. Pour chaque échantillon, la simulation de digestion haute a été menée en parallèle sur cinq sous-échantillons, et les fractions IF et SF de ces cinq digestions indépendantes ont été rebroyées, pesées et rassemblées pour le dosage des fibres alimentaires totales (TDF) et les mesures de fermentation microbienne.

### 1.4. Dosage des fibres alimentaires totales

Les fibres alimentaires totales sont déterminées sur des réplicas d'échantillons de matières sèches (Prosky et al., 2020). Brièvement, les échantillons subissent une digestion de l'amylase (30 min, 100°C), puis de protéase (30 min, 60°C) et d'amyloglucosidase (30 min, 60°C). Le résidu de la précipitation faite avec 4 volumes d'éthanol est filtré, lavé avec de l'éthanol à 78%, de l'éthanol à 95% et de l'acétone, puis séché et pesé. Un premier réplica est analysé pour les protéines (méthode de Dumas) et le second est calciné à 550°C pour déterminer les cendres brutes. Le TDF, exprimé en pourcentage de matière sèche, correspond au poids du résidu filtré et séché dont la teneur de protéines et cendres brutes a été soustraite.

### 1.5. Fermentation microbienne

Les échantillons de fèces ont été décongelés à température ambiante, un procédé qui a peu d'impact sur la fermentation (Deschamps et al., 2020). Ils ont été pesés dans la chambre anaérobie afin de constituer 10 mL de suspension fécale à 10% pour chaque échantillon. Ensuite 1,6 mL ont été répartis dans des tubes Eppendorf de 2 mL pré-remplis avec les fibres testées : aliment brut broyé, IF de l'aliment, amidon de maïs ou beta-glucanes d'avoine. Des tubes sans ajout de fibres ont été ajoutés afin d'estimer la fermentation basale qui a lieu à partir des substrats disponibles dans le contenu digestif. Un tube a été renversé accidentellement. La fermentation a eu lieu dans un tampon PBS 10 mM équilibré dans une chambre anaérobie contenant 10 ppm d'oxygène, 5% de $CO_2$, 1,5% de $H_2$ et 93,5% de $N_2$.

### 1.6. Caractérisation de la biomasse microbienne

A la fin de l'incubation, l'ADN de 200 µL de biomasse a été extrait avec le kit Zymo ZR 96 et la région V3-V4 du gène de l'ARN ribosomal 16S a été séquencée (Zemb et al., 2024). L'analyse des ASV (Amplicon Sequence Variants) a été réalisé sous R avec DADA2 comme pour les données publiées précédemment (Aliakbari et al., 2021). La table de 1 326 ASVs a été simplifiée en 148 genres par la fonction AssignTaxonomy sur la base de données silva_nr_v138_train_set (Quast et al., 2013).

### 1.7. Mesure des métabolites fermentaires

Pour l'analyse des acides gras à chaîne courte, les tubes Eppendorf contenant 1,8 mL du produit d'incubation ont été centrifugés 20 min à 8000 g, puis 1,5 mL de surnageant a été mélangé avec 300 µL d'acide métaphosphorique à 25 %. Les surnageants ont été collectés et conservés à -20°C. La composition des acides gras à chaine courte a été déterminée par chromatographie en phase gazeuse selon Playne (Playne, 1985). L'analyse du lactate, du succinate et du glucose a été réalisé à partir de 200 µL de surnageant acidifié passés par chromatographie de haute performance à phase liquide (Vanquish, ThermoFisher) sur une colonne Rezex ROA H+ (Phenomenex) et une élution isocratique avec H2SO4 0,5 mM à 0,5 mL/min. Les composés sont déterminés à l'aide d'un détecteur ultra-violet à 205 nm et d'un détecteur à indice de réfraction. Les productions spécifiques de chaque source de fibres ont été évaluées en soustrayant la concentration de chaque métabolite des incubations contrôles aux concentrations observées dans les incubations avec fibres.

**1.8. Statistiques**

Toutes les analyses statistiques ont été réalisées à l'aide du logiciel R (version 4.3.0). Les concentrations de chaque acide gras à chaine courte ont été analysées par un test de Wilcoxon pour déterminer les différences entre fermentations de porcs CMJR– et de porcs CMJR+. Les différences d'abondances de genres microbiens ont été comparées par un modèle linéaire avec correction de Benjamini-Hochberg. Les analyses de corrélations entre production d'acides gras à chaine courte et performances se basent sur des calculs de corrélations de Spearman.

**2. RESULTATS ET DISCUSSION**

**2.1. Différences de fonctions inférées entre lignées**

Les fonctions prédites par PICRUST2 sur 588 animaux apparaissent dans neuf voies métaboliques qui sont associées positivement avec la valeur de CMJR des porcs, dans un modèle corrigeant pour l'effet lignée (Tableau 1).

|  | Voie métabolique inférée | log(f) | q-val |
|---|---|---|---|
| ko03010 | Ribosome | 4,8 | 3,2E-04 |
| ko01240 | Biosynthèse de cofacteurs | 3,0 | 2,6E-02 |
| ko01230 | Biosynthèse des acides amines | 4,5 | 3,2E-04 |
| ko01200 | Métabolisme du carbone | 3,6 | 6,1E-03 |
| Ko00970 | Biosynthèse de l'aminoacyl-tRNA | 3,1 | 2,6E-02 |
| ko00680 | Métabolisme du méthane | 2,6 | 6,0E-02 |
| ko00550 | Biosynthèse des peptidoglycans | 3,6 | 7,5E-03 |
| ko00270 | Métabolisme de la cystéine et de la méthionine | 3,6 | 6,1E-03 |
| ko00020 | Cycle TCA | 3,0 | 2,6E-02 |

**Tableau 1** – Voies métaboliques microbiennes associées à la valeur de CMJR Identifiant de voie métabolique

De façon intéressante, l'enrichissement des voies de biosynthèse des acides aminés et de cofacteur suggère une activité du microbiote différente. L'enrichissement de la voie de biosynthèse de l'aminoacyl-tRNA semble associée à une production élevée de propionate dans d'autres études chez le porcelet. En effet, la manipulation du microbiote de porcelet par transplantation fécale ou par une communauté de microbes cultivables a abouti à une hausse de cette voie de façon concomitante avec la hausse du propionate (Rahman et al., 2023). Nous avons donc décidé d'étudier plus en détail la production de propionate par des fermentations in vitro simulant la digestion.

**2.2. Différences de fermentation mesurées**

Le pH final des incubations se situe entre 5,5 et 6. Ces valeurs sont pertinentes physiologiquement puisque le pH dans le tube digestif du porc se situe entre 5,5 et 7,5 (Henze et al., 2021), ce qui montre que le pouvoir tampon du système in vitro est suffisant par rapport à la quantité de fibres fermentées.

Les fermentations à partir de la fraction IF de l'aliment, qui représente 28 ± 8% dans cette étude, montrent que les porcs CMJR− produisent plus d'acétate (+15%) et de propionate (+56%) (Tableau 2), ce qui génère plus d'acides gras à courte chaine. Bien que la quantité de valérate ne soit pas en moyenne significativement différente entre les lignées, la lignée CMJR− produit plus fréquemment du valérate à hauteur d'au moins 30 mM par gramme d'aliment, puisque seuls 2/23 porcs CMJR− produisent moins que 30 mM de valérate, contre 10/24 porcs CMJR+ (Figure 1, P < 0,009 pour un test de Chi² de répartition des animaux dans ces deux catégories).

| Acide | | CMJR+ | CMJR− | Δ | P |
|---|---|---|---|---|---|
| Fraction insoluble (*65% TDF, 5,3 % prot*) | Acetate | 400 | 463 | +15% | 0,005 |
| | Propionate | 128 | 200 | +56% | 0,018 |
| | Butyrate | 218 | | NS | |
| | Valérate | 65 | | NS | |
| | Glucose | 1/24 | 5/23 | NA | 0,01 |
| | Formate | 0,9 | | NS | |
| | Succinate | 1,3 | | NS | |
| | Lactate | 0 | | NS | |
| Aliment brut (*17% TDF, 7,6% prot*) | Acetate | 263 | | NS | |
| | Propionate | -124 | | NS | |
| | Butyrate | -74 | | NS | |
| | Valérate | -74 | | NS | |
| β-glucanes | Acetate | 222 | | NS | |
| | Propionate | -24 | | NS | |
| | Butyrate | -64 | | NS | |
| | Valérate | -54 | | NS | |
| Amidon (*73% amylopectin and 27% amylose*) | Acetate | 341 | 468 | +37% | 0,02 |
| | Propionate | 61 | | NS | |
| | Butyrate | 179 | | NS | |
| | Valérate | 26 | | NS | |

**Tableau 2** – Production de métabolites après 72 h d'incubation avec diverses sources de fibres (en mM, sauf glucose) chez les porcs des lignées CMJR− et CMJR+ (*une production négative correspond à une concentration supérieure dans les contrôles sans substrat*).

Les consortia microbiens produisent des niveaux faibles mais significatifs de succinate à partir de la fraction IF de l'aliment (*P* < 0,04) mais de façon similaire entre les deux lignées (*P* > 0,40). Le lactate est détecté dans les échantillons mais ne diffère pas des fermentations contrôles sans IF, d'où une production nette nulle. La production de glucose tend à être détectée plus fréquemment dans les porcs CMJR− (6/24) que dans les porcs CMJR+ (1/23, *P* = 0,08 pour un test de Chi² de répartition des animaux dans ces deux catégories). Ces résultats suggèrent que les consortia microbiens provenant de porcs CMJR− fermentent plus efficacement la fraction IF de l'aliment par la voie du succinate.

A partir de l'aliment brut broyé (Tableau 2), aucune différence significative de fermentation n'est détectée. Le niveau plus faible de production d'acétate et les productions négligeables des autres acides gras à courte chaine suggèrent que les oses sont moins accessibles aux microbes si l'aliment n'est pas préalablement soumis aux enzymes digestives, ce qui a déjà été observé à partir de la production de gaz fermentaires avec et sans traitement de pepsine-pancréatine (Bindelle *et al.*, 2007).

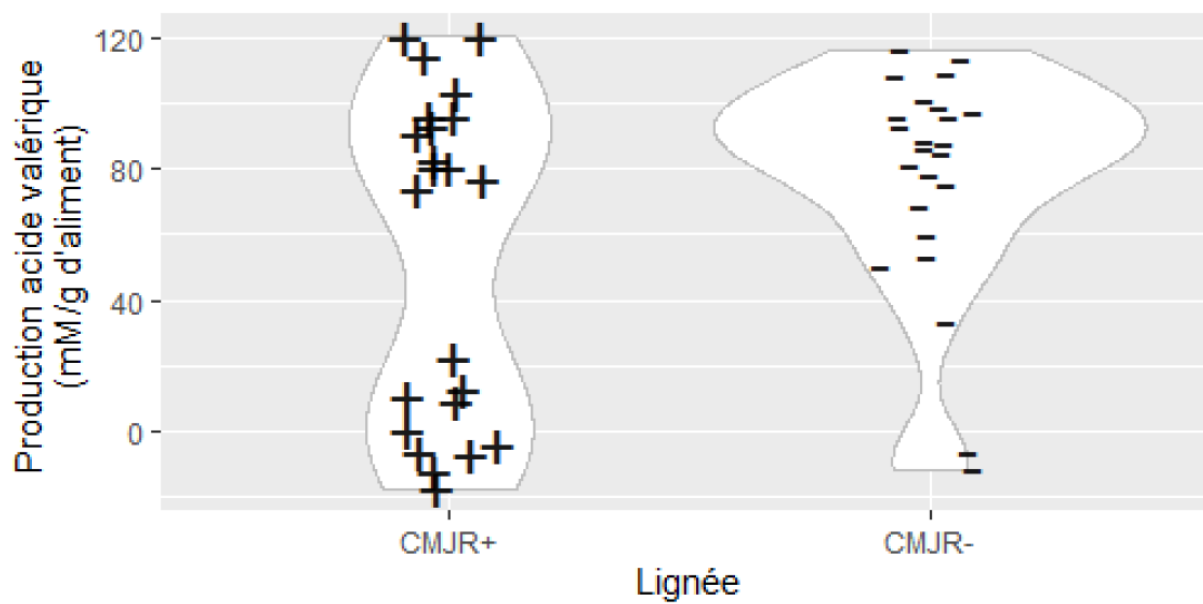

**Figure 1** – Production de valérate dans les lignées CMJR+ et CMJR– à partir de la fraction IF

A partir d'amidon résistant (Tableau 2), les porcs CMJR–produisent 37% plus d'acétate, ce qui suggère que les liaisons α (1→4) et α (1→6) sont attaquées plus efficacement par les microbiotes de porcs CMJR–. Le manque de différence de fermentation sur les β-glucanes d'avoine suggère que les oses linéaires reliés par des liaisons β (1→3) et β (1→4) sont trop facilement accessibles pour permettre aux consortia des porcs CMJR– et CMJR+ d'illustrer leur différence. Il est cependant surprenant que les β-glucanes ne génèrent pas de butyrate comme observé précédemment (Bai et al., 2021).

Lorsqu'aucune fibre n'est ajoutée (Tableau 2), les suspensions fécales présentent tout de même une fermentation basale puisque la concentration d'acétate, de propionate et de butyrate des tubes contrôles (sans ajout de fibres) sont respectivement de 8, 4 et 2 mM après 72 h d'incubation. Ces valeurs de fermentations basales ne dépendent pas de la lignée et sont de l'ordre de grandeur attendu puisque l'analyse d'échantillons de contenus intestinaux collectés sur des cadavres de porcs a montré des concentrations d'acétate de 52 mM, de propionate de 20 mM et de butyrate de 6 mM (Cummings et al., 1987), ce qui est proche des valeurs obtenues ici si on tient compte de la dilution à 10% dans le tampon PBS. Ces contrôles de fermentations basales ont également une faible teneur en acide isobutyrique (0,6 mM) et isovalérique (0,9 mM) qui dépendent de la lignée. Le glucose et le lactate ne sont pas détectés de façon significative dans les fermentations contrôles. Pris ensemble, ces résultats indiquent que les lignées produisent les mêmes quantités basales d'acétate, de propionate et de butyrate à partir des substrats disponibles dans le contenu intestinal mais pourraient être légèrement différentes en termes de fermentation de la valine et de la leucine, qui produisent l'isobutyrate et l'isovalérate.

Ces résultats suggèrent que les consortia de porcs CMJR–fermentent la fraction IF par la voie du succinate, et que les régimes à fortes teneur en fibres favorisent davantage la mise en évidence des différences fermentaires. Ces résultats sont cohérents avec l'observation que chez les porcs nourris avec un régime riche en fibres, les différences d'efficacité alimentaire étaient associées à 17 unités taxonomiques opérationnelles (OTU) chez les mâles (P = 0,018) et à 7 OTU chez les femelles (P = 0,010), avec 3 OTU en commun pour les deux sexes (Verschuren et al., 2018).

## 2.3. Microbes responsables des fermentations à partir de la fraction insoluble

L'analyse de la biomasse de chaque incubation par séquençage du gène ribosomal 16S a permis d'identifier les microbes fermentant la fraction IF de l'aliment à partir des 148 genres microbiens identifiés (Tableau 3). Ces microbes appartiennent aux familles des Veillonellaceae (*Dialister* et *Mitsukella*), des Prevotellaceae (*Prevotella*), des Lactobacillaceae (*Lactobacillus*), des Lachnospiraceae (*Shuttleworthia*) ou des Bifidobacteriaceae (*Bifidobacterium*). *Prevotella* représente la majorité des microbes fermentaires, ce qui est en accord avec les observations faites pour les microbiotes humains de l'entérotype '*Prevotella*', où les fermentations d'oligosaccharides et d'arabinoxylanes sont principalement assurées par une seule OTU de *Prevotella* (Chen *et al.*, 2017). Dans nos données, l'abondance de *Prevotella* change entre les groupes, suggérant une réponse de fermentation variable. Cependant, l'abondance de *Prevotella* à la fin de la fermentation ne diffère pas entre porcs CMJR− et CMJR+. Il semblerait donc que *Prevotella* assure l'essentiel de la fermentation dans les deux lignées, mais que cela n'explique pas la plus grande efficacité des CMJR−, où il est moins abondant. *Bifidobacterium*, *Dialister* et *Lactobacillus* semblent expliquer la différence de fermentation entre CMJR+ et CMJR−.

|  | CMJR+ | | CMJR- | | | | |
|---|---|---|---|---|---|---|---|
| **Genre** | **CTL** | **IF** | **CTL** | **IF** | **Effet IF** | **Effet CMJR** | **Effet bande** |
| **Prevotella** | 1,3 | 42,2 | 1,1 | 55,1 | *** | NS | * |
| **Lactobacillus** | 15,1 | 37,5 | 4,0 | 19,9 | *** | *** | ** |
| **Shuttleworthia** | 3,3 | 4,6 | 1,8 | 3,2 | ** | NS | * |
| **Dialister** | 0,7 | 2,2 | 0,6 | 1,2 | *** | * | * |
| **Bifidobacterium** | 0,1 | 1,7 | 0,0 | 0,4 | *** | ** | NS |
| **Mitsuokella** | 0,1 | 0,2 | 0,0 | 0,1 | *** | NS | NS |

**Tableau 3** – Abondance (pourcentage de la biomasse microbienne à la fin de l'incubation) des microbes fermentant la fraction IF de l'aliment après 72h.

## 3. LIEN ENTRE FERMENTATIONS MICROBIENNES ET PERFORMANCES *IN VIVO*

### 3.1. Lien entre fermentation in vitro et métabolisme lipidique

Chez les porcs à faible CMJR, la corrélation la plus forte entre les produits de fermentation et les performances est celle entre la quantité de propionate produite par fermentation in vitro de la fraction IF et l'épaisseur du lard dorsal (Corr$_{Spearman}$=0,8, Figure 2). La production d'acétate à partir de la fraction IF *in vitro* est modérément corrélée à la consommation moyenne journalière (Corr$_{Spearman}$= 0,5).

La corrélation entre le propionate produit et l'épaisseur de lard dorsal suggère un lien avec le métabolisme adipeux, tel que décrit dans la littérature. En effet, des porcs ayant reçu des infusions cæcales de propionate (36 mmol/kg$_{0,75}$ par jour) ont vu leur le cholestérol sérique total augmenté de 15% et le cholestérol LDL de 15% (Beaulieu, McBurney, 1992). Lorsqu'un mélange d'acides acétique, propionique et butyrique est infusé intracæcalement chez des porcs, la majeure partie de l'énergie infusée en acides gras à courte chaine était retenue sous forme de graisse corporelle estimée d'après les données de chambres respiratoires (Jørgensen *et al.*, 1997). Cet effet découle probablement de l'acétate et du propionate qui favorise l'accumulation de lipides dans les adipocytes via l'activation du récepteur FFA2/GPR43 inhibant la lipolyse (Hong *et al.*, 2005). Il semble donc logique que la production de propionate soit associée à plus de lard dorsal comme observé chez les porcs CMJR−.

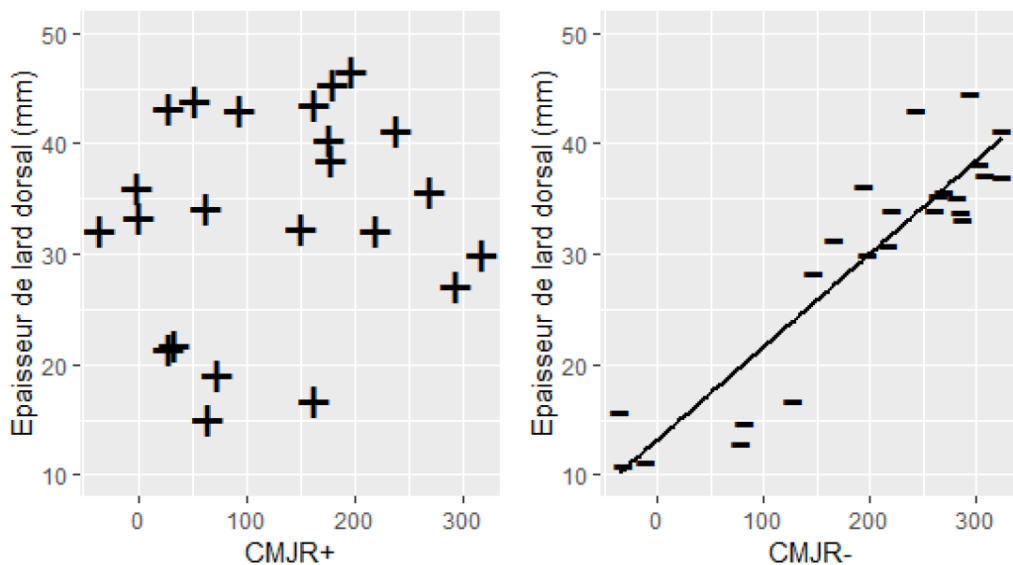

**Figure 2** – Distribution conjointe de la production de propionate *in vitro* obtenue à partir de la fraction IF (mM/g IF) et de l'épaisseur de lard dorsal *in vivo* chez les porcs CMJR+ (à gauche) et CMJR− (à droite)

**3.2. Lien entre fermentation et efficacité alimentaire**

Chez les porcs CMJR+, la production de butyrate est légèrement mais significativement corrélée à l'indice de consommation (Corr$_{Spearman}$=0,44, Figure 3), ce qui ne se trouve pas chez les porcs CMJR−. Toutes les autres corrélations ne sont pas significatives.

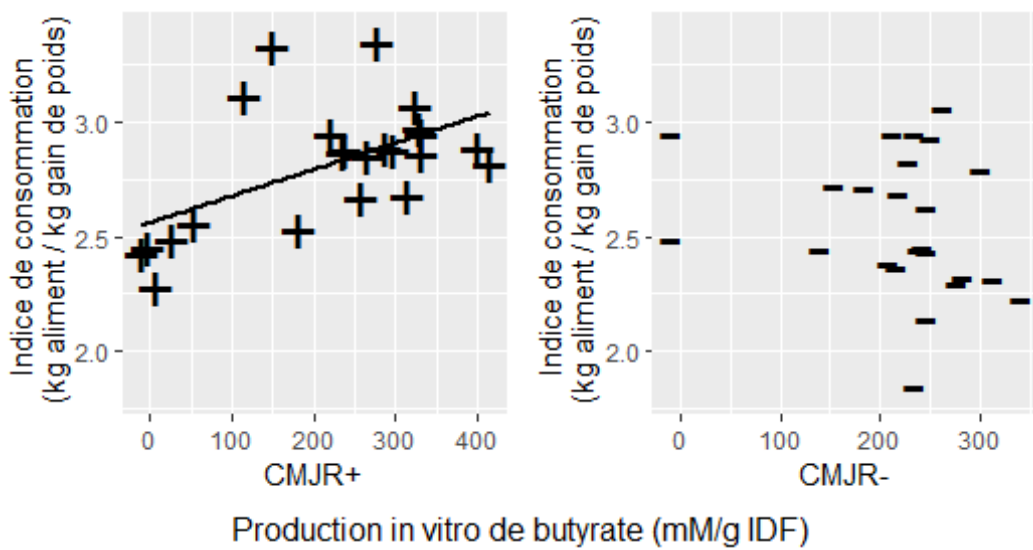

Production in vitro de butyrate (mM/g IDF)

**Figure 3** - Distribution conjointe de la production de butyrate *in vitro* obtenue à partir de la fraction IF et de l'indice de consommation *in vivo* chez les porcs CMJR+

**3.3. Avantages et limitations du modèle *in vitro***

Il est crucial de mettre au point un système in vitro qui soit suffisamment proche des conditions in vivo pour garantir la pertinence des conclusions.

La mesure des profils fermentaires montre que les échantillons fécaux des porcs CMJR– génèrent plus de propionate à partir de la fraction IF de l'aliment. Ce point n'est pas complètement cohérent avec les inférences réalisées avec PICRUST2 sur la base des informations génomiques, ce qui souligne le potentiel de la validation fonctionnelle in vitro pour consolider les inférences réalisées in silico.

Cependant, certaines limites existent à cette approche. Notre protocole de prédigestion est identique pour les porcs CMJR– et CMJR+. Notre modèle ne prend donc pas en compte les différences entre lignées qui peuvent avoir lieu dans la partie haute du tube digestif, comme par exemple le transporteur de folate qui est surexprimé dans le duodénum des porcs CMJR– (Devailly et al., 2023), et aboutir à des bols alimentaires résultant de traitements légèrement différents à l'entrée du colon entre porcs CMJR– et CMJR+.

Une autre limitation de notre modèle est l'absence d'absorption des acides gras à courte chaine, ce qui entraîne leur accumulation à des concentrations jusqu'à 10 fois plus élevées que celles observées in vivo. Par exemple, l'acétate atteint des concentrations de 80 mM avant les calculs de normalisation par gramme d'aliment, ce qui pourrait impacter certaines voies microbiennes. La croissance d'E. coli sur glucose est diminuée lorsqu'on ajoute 128 mM d'acétate (Pinhal et al., 2019). Dans notre cas, la résistance aux concentrations élevées d'acides à courte chaine pourrait en partie expliquer les dominances de 50% Prevotella et de 27% de Lactobacillus à la fin de l'incubation. Ces dominances ne sont pas aussi marquées in vivo, où ces espèces atteignent 15 à 20%, ce qui montre que certains mécanismes de régulation des populations sont absents de notre modèle axé sur la fermentation microbienne.

Parmi les mécanismes expliquant la faible diversité des phylotypes en croissance, le manque de diversité des substrats est une piste probable. Par exemple, notre modèle ne prend pas en compte les mucines de l'hôte, qui sont un substrat potentiel. En effet, l'ajout de mucines de l'hôte peut augmenter légèrement la quantité d'acétate produit à partir de cellulose et favoriser les Lachnospiraceae (Tran et al., 2016). En outre, notre modèle considère que la fraction solubilisée par les enzymes digestives est totalement assimilée par l'intestin grêle, ce qui n'est pas le cas. D'ailleurs, certains auteurs recommandent l'ajout de 10% de fraction soluble (Pérez-Burillo et al., 2021). D'une manière générale, cet apport de fraction soluble riche en protéines de l'aliment pourraient faire l'objet d'une compétition majeure pour les microbes intestinaux (Holmes et al., 2017). La quantité et la qualité exactes de molécules solubles qui arrivent au colon est difficile à estimer puisque certains sucres solubles, comme le tagatose, arrivent au colon à 75% (Laerke, Jensen, 1999), tandis que le glucose est complètement absorbé par l'intestin grêle (Hall, Guyton, 2011).

**CONCLUSION**

La production de propionate dans notre modèle in vitro est significativement corrélée à l'épaisseur de gras dorsal mesurée in vivo dans la lignée CMJR-. Ceci suggère que le modèle in vitro parvient à capter certaines différences fonctionnelles entre les communautés microbiennes des porcs CMJR– et CMJR+ qui n'étaient pas bien prédites par l'inférence fonctionnelle à partir des données de séquençage, montrant la complémentarité entre les outils. De plus, cela suggère qu'il est possible d'utiliser le modèle in vitro pour évaluer le potentiel fermentaire des aliments et d'optimiser la composition de l'aliment à la génétique du porc.

**REMERCIEMENTS**